\documentclass[11pt,a4paper]{article}
\usepackage{amsfonts}
\usepackage{amsmath,amssymb}
\usepackage{epsfig,graphicx}

\input{tcilatex}
\begin{document}

\bigskip

\bigskip

\begin{center}
{\LARGE Gauge Symmetries Emerging from Extra Dimensions }

{\LARGE \bigskip \bigskip }

\textbf{J.L.~Chkareuli and Z. Kepuladze}

$^{1}$\textit{Center for Elementary Particle Physics, Ilia State University,
0162 Tbilisi, Georgia\ \vspace{0pt}\\[0pt]
}

$^{2}$\textit{E. Andronikashvili} \textit{Institute of Physics, 0177
Tbilisi, Georgia\ }

\textit{\bigskip }

\bigskip

\bigskip

\bigskip

\bigskip

\textbf{Abstract}

\bigskip \bigskip
\end{center}

We argue that extra dimensions with a properly chosen compactification
scheme could be a natural source for emergent gauge symmetries. Actually,
some proposed vector field potential terms or polynomial vector field
constraints introduced in five-dimensional Abelian and non-Abelian gauge
theory is shown to smoothly lead to spontaneous violation of an underlying
5D spacetime symmetry and generate pseudo-Goldstone vector modes as
conventional 4D gauge boson candidates. As a special signature, there
appear, apart from conventional gauge couplings, some properly suppressed
direct multi-photon (multi-boson, in general) interactions in emergent QED
and Yang-Mills theories whose observation could shed light on their
high-dimensional nature. Moreover, in emergent Yang-Mills theories an
internal symmetry $G$ also occurs spontaneously broken to its diagonal
subgroups once 5D Lorentz violation happens. This breaking origins from the
extra vector field components playing a role of some adjoint scalar field
multiplet in the 4D spacetime. So, one naturally has the Higgs effect
without a specially introduced scalar field multiplet. Remarkably, when
being applied to Grand Unified Theories this results in a fact that the
emergent GUTs generically appear broken down to the Standard Model just at
the 5D Lorentz violation scale $\mathrm{M.}$ \bigskip\ \bigskip

\thispagestyle{empty}\newpage

\section{Introduction and overview}

A significant progress in understanding of the spontaneously broken internal
symmetries with accompanying massless scalar Goldstone modes \cite{NJL}
allows one to think that spacetime symmetries, and first of all Lorentz
invariance, could also be spontaneously broken so as to generate massless
vector and tensor Goldstone modes associated with photons, gravitons and
other gauge fields. This has attracted a considerable interest over the last
fifty years in many different contexts which could be basically classified
as the composite models \cite{bjorken, ph, eg, suz}, constraint-based models 
\cite{nambu} and potential-based models \cite{ks} (for some later
developments see \cite{cfn, kraus, jen, bluhm, kos, car, cjt, cfn2, c3}). We
give below some short formulation of them to make clearer the aims of the
present paper.

\subsection{Composite models}

Composite models are based on the four-fermi (or multi-fermi in general)
interaction where the photon and other gauge fields may appear as a
fermion-antifermion pair composite state in complete analogy with massless
composite scalar fields (identified with pions) in the original
Nambu-Jona-Lazinio model \cite{NJL}. This old idea is better expressed
nowadays in terms of effective field theory where the standard \ QED
Lagrangian is readily obtained through the corresponding loop radiative
effects due to $N$ fermion species involved \cite{kraus, jen}. One could
think, however, that composite models contain too many prerequisites and
complications related to the large number of basic fermion species involved,
their proper arrangement, non-renormalizability of the fundamental
multi-fermi Lagrangian,\ instability under radiative corrections, and so on
indefinitely. This approach contains in fact a cumbersome invisible sector
which induces the effective emergent theory. A natural question then arises
whether one could directly work in the effective vector field theory instead
thus having spontaneous Lorentz invariance violation (SLIV) from the outset.

\subsection{Potential-based models}

Actually, one could start with a conventional QED type Lagrangian extended
by an arbitrary vector field potential energy terms which explicitly break
gauge invariance. For a minimal potential containing bilinear and quartic
vector field terms one comes to the Lagrangian%
\begin{equation}
L_{V}=L_{QED}-\frac{\lambda }{4}\left( A_{\mu }A^{\mu }-n^{2}M^{2}\right)
^{2}  \label{lag2}
\end{equation}%
where the mass parameter $n^{2}M^{2}$ stands for the proposed SLIV scale,
while $n_{\mu }$ is a properly-oriented unit Lorentz vector, $n^{2}=n_{\mu
}n^{\mu }=\pm 1$. This partially gauge invariant model being sometimes
referred to as the \textquotedblleft bumblebee\textquotedblright\ model \cite%
{ks} (see also \cite{bluhm} and references therein) means in fact that the
vector field $A_{\mu }$ develops a constant background value 
\begin{equation}
<A_{\mu }>\text{ }=n_{\mu }M\text{ }  \label{vev1}
\end{equation}%
and Lorentz symmetry $SO(1,3)$ breaks down at the proposed SLIV scale $M$ to 
$SO(3)$ or $SO(1,2)$ depending on whether $n_{\mu }$ is time-like ($n^{2}=+1$%
) or space-like ($n^{2}=-1$). Expanding the vector field around this vacuum
configuration, 
\begin{equation}
A_{\mu }(x)=a_{\mu }(x)+n_{\mu }(M+H),\text{ \ \ }n_{\mu }a_{\mu }=0
\label{ex}
\end{equation}%
one finds that the $a_{\mu }$ field components, which are orthogonal to the
Lorentz violating direction $n_{\mu }$, describe a massless vector
Nambu-Goldstone boson, while the $H$ field corresponds to a massive Higgs
mode away from the potential minimum. Due to the presence of this mode the
model may in principle lead to some physical Lorentz violation in terms of
the properly deformed dispersion relations for photon and matter fields
involved that appear from the corresponding radiative corrections to their
kinetic terms \cite{kraus}. However, as was argued in \cite{vru}, a
bumblebee-like model appears generally unstable\footnote{
Apart from the instability, the potential-based models were shown \cite{aki}
to be obstructed from having a consistent ultraviolet completion, whereas
the most of viable effective theories possess such a completion.}, its
Hamiltonian is not bounded from below beyond the constrained phase space
determined by the nonlinear condition 
\begin{equation}
A_{\mu }A^{\mu }=n^{2}M^{2}\text{ .}  \label{const}
\end{equation}%
With this condition imposed, the massive Higgs mode never appears, the
Hamiltonian is positive, and the model is physically equivalent to the
nonlinear constraint-based QED, which now we briefly consider.

\subsection{Constraint-based models}

This class of models starts directly with the nonlinearly realized Lorentz
symmetry for underlying vector field (or vector field multiplet) through the
"length-fixing" constraint (\ref{const}) implemented into conventional gauge
invariant theories, both Abelian and non-Abelian ones. This constraint in
itself was first studied in the QED framework by Nambu quite a long time ago 
\cite{nambu}, and in a general context (including loop corrections \cite{az}%
, massive QED framework \cite{kep}, non-Abelian \cite{jej, cj, wiki} and
supersymmetric \cite{c} extensions) in the last decade. The constraint-based
models show that, in contrast to the spontaneous violation of internal
symmetries, spontaneous Lorentz violation producing vector Goldstone bosons
seems not to necessarily imply physical breakdown of Lorentz invariance.
Rather, when appearing in a gauge theory framework, this may eventually
result in a noncovariant gauge choice in an otherwise gauge invariant and
Lorentz invariant theory.

Rather than impose by postulate, the constraint (\ref{const}) may be
implemented into the standard QED Lagrangian $L_{QED}$ through the invariant
Lagrange multiplier term

\begin{equation}
L=L_{QED}-\frac{\mathrm{\lambda }}{2}\left( A_{\mu }A^{\mu
}-n^{2}M^{2}\right) \text{ , \ }n^{2}=n_{\mu }n^{\mu }=\pm 1  \label{lag}
\end{equation}%
provided that initial values for all fields (and their momenta) involved are
chosen so as to restrict the phase space to values with a vanishing
multiplier function \textrm{$\lambda $}$(x)$, \textrm{$\lambda $ }$=0$.
Actually, due to an automatic conservation of the matter current in QED an
initial value $\mathrm{\lambda }=0$ will then remain for all time so that
the Lagrange multiplier field \textrm{$\lambda $} never enters in the
physical equations of motions for what follows\footnote{%
Interestingly, this solution with the Lagrange multiplier field $\mathrm{%
\lambda }(x)$ being vanished can technically be realized by introducing in
the Lagrangian (\ref{lag}) an additional Lagrange multiplier term of the
type $\mathrm{\xi \lambda }^{2}$, where $\mathrm{\xi }(x)$ is a new
multiplier field. One can now easily confirm that a variation of the
modified Lagrangia $L+$ $\mathrm{\xi \lambda }^{2}$ with respect to the $%
\mathrm{\xi }$ field leads to the condition $\mathrm{\lambda }=0$, whereas a
variation with respect to the basic multiplier field $\mathrm{\lambda }$
preserves the vector field constraint (\ref{const}).}. It is worth noting
that, though the Lagrange multiplier term formally breaks gauge invariance
in the Lagrangian (\ref{lag}), this breaking is in fact reduced to the
nonlinear gauge choice (\ref{const}). On the other hand, since gauge
invariance is no longer generically assumed, it seems that the vector field
constraint (\ref{const}) might be implemented into the general vector field
theory (\ref{lag2}) rather than the gauge invariant QED in (\ref{lag}). The
point is, however, that both theories are equivalent once the constraint (%
\ref{const}) holds. Indeed, due to a simple structure of vector field
polynomial in (\ref{lag2}), they lead to practically the same equations of
motion in both cases.

The constraint (\ref{const}) is in fact very similar to the constraint
appearing in the nonlinear $\sigma $-model for pions \cite{GLA}. It means,
in essence, that the vector field $A_{\mu }$ develops some constant
background value, $<A_{\mu }(x)>$ $=n_{\mu }M$, and has a special
"Higgsless" expansion around vacuum configuration%
\begin{equation}
A_{\mu }=a_{\mu }+n_{\mu }\sqrt{M^{2}-n^{2}a^{2}}\text{ , \ }n_{\mu }a_{\mu
}=0\text{ \ }(a^{2}\equiv a_{\mu }a^{\mu })  \label{gol}
\end{equation}%
so that Lorentz symmetry formally breaks down, depending on a particular,
time-like or space-like, nature of SLIV mentioned above. The point is,
however, that, in sharp contrast to the nonlinear $\sigma $-model for pions,
the nonlinear QED theory ensures that all the physical Lorentz violating
effects strictly cancel out among themselves (as was explicitly shown both
in the tree \cite{nambu} and one-loop \cite{az} approximations), due to the
starting gauge invariance involved. The non-covariant gauge choice for
vector Goldstone bosons shown in (\ref{gol}) appears as the only response of
the theory to SLIV.

So to conclude, although it may sound somewhat counterintuitive, one may
separate these two aspects: generation of vector Goldstone bosons and
physical Lorentz violation. When such a spontaneous violation occurs in the
gauge invariant vector field system, this system generates massless
Goldstone modes paying for that just gauge degrees of freedom and leaving
the physical ones untouched. As to an observational evidence in favor of
emergent theories the only way for SLIV to cause physical Lorentz violation
would appear only if gauge invariance in these theories were really broken
rather than merely constrained by some gauge condition. Such a violation of
gauge invariance could provide the potential-based model considered above or
some extension of the constraint-based model with high-dimension operators
induced by gravity at very small distances \cite{par}. However, in any case,
if we are primarily interested in the vector Goldstone generation rather
than physical Lorentz violation, it seems more relevant to work in the
framework of the constraint-based models rather than in the largely
contradictory potential-based ones. We will follow this strategy for the
rest of the paper.

\subsection{Models with extra spacetime dimensions}

Now, after this brief sketch of valuable SLIV models one can\ see that all
of them only suggest a non-covariant description of vector Goldstone bosons
where one of vector field spacetime component $A_{\mu }$ ($\mu =0,1,2,3)$ is
"higgsified" (\ref{ex}) or constrained (\ref{gol}). It is rather clear that
the only way to produce the vector Goldstone bosons in the fully Lorentz
covariant way, both in the potential-based and the constraint-based models,
would be to enlarge the existing Minkowski spacetime to higher dimensions.
Particularly, the spontaneous breakdown of the "five-dimensional Lorentz
symmetry" to the ordinary one, $SO(1,4)\rightarrow SO(1,3),$ could generate
a conventional four-dimensional vector Goldstone vector field $A_{\mu }$ ($%
\mu =0,1,2,3$) that was first argued quite a long ago \cite{cfn, li, rabi},
though has not been yet worked out in significant detail. Remarkably, the
requirement for a fully covariant description of vector Goldstone fields may
have, as we will see later, far going consequences for emergent gauge
theories. Actually, in contrast to the above mentioned 4D models with the
hidden SLIV, now due to the proposed compactification scheme\ to physical
four dimensions, the starting 5D gauge invariance in these theories appears
broken. This does not allow to gauge away from them some possible
observational evidence in favor of their emergent nature.

One could try to implement the high-dimensional SLIV\ program into the brane
models \cite{add} with our physical world assumed to be located on a
three-dimensional brane embedded in the high-dimensional bulk. However, a
serious problem for such theories seems to be how to achieve the
localization of emergent gauge fields on the flat brane associated with our
world \cite{rub}. In this connection, more attractive possibility seems to
be related to a class of extra-dimensional models known as Universal Extra
Dimensions (UED) \cite{app, pil, hoop}. In them the Standard Model fields
(or, at least, some essential part of them) are free to propagate through
all of the dimensions of space, rather than being confined to our physical
spacetime as they typically are in the brane models. Naturally, the UED
models look more similar to the original Kaluza-Klein \ proposal than
somewhat more sophisticated brane model scenarios \cite{add, rs}.
Phenomenologically, the UED models with the KK parity involved considerably
relaxes the constraints from electroweak precision data, allowing for much
lower scales of compactification $M_{KK}$ $=1/R$ (even up to a few $TeV$
order scale). Another important aspect related to them appears in its
ability to provide a natural candidate for the dark matter in the universe.
In particular, the lightest KK state can be stable and produced in the early
universe with an abundance similar to that of the measured dark matter
density. One more attractive feature seems to be that the UEDs, as was
mentioned above, could also be a natural source for vector Goldstone bosons
associated with photons and other gauge fields, particularly if one proceeds
in the five-dimensional UED framework.

\subsection{The present paper}

We argue that extra dimensions with a properly chosen compactification
scheme could be a natural source for emergent gauge symmetries. We start
with a simple QED type theory with the SLIV\ in five-dimensional (5D)
spacetime. This 5D SLIV could appear due to some vector field constraint
being a high-dimensional analog of the constraint considered above (\ref%
{const}), as is argued in section 2. This lead to the spontaneous violation
of the 5D Lorentz symmetry at some high scale $\mathrm{M}$ that proposedly
goes along with a compactification of the 5D spacetime down to physical four
dimensions at the comparable scale $M_{KK}$. This is in fact the symmetrical
orbifold compactification $S_{1}/Z_{2}$ under which all spacetime components
of the 5D vector field $A_{\overline{\mu }}$ ($\overline{\mu }=0,1,2,3;5$)
are taken to be even. The important point is that such a compactification,
which breaks the starting 5D gauge invariance to a conventional 4D gauge
invariance for the vector field ground modes $A_{\mu }^{0}$\ ($\mu =0,1,2,3$%
) may significantly contribute into the physical processes involved. As a
special signature, there appear, apart from conventional gauge couplings,
some properly suppressed direct multi-photon (multi-boson, in general)
interactions in emergent QED and Yang-Mills theories. This means that they
actually possess only a partial gauge invariance whose observation could
shed light on their high-dimensional nature. In section 3 we turn to the
Yang-Mills theories where not only spacetime symmetry but also internal
symmetry appears spontaneously broken once the 5D SLIV happens. Remarkably,
this breaking looks like the breaking that is usually induced by an
appropriate adjoint scalar field multiplet incorporated into the vector
field theory. Now this breaking origins from extra vector field components.
Therefore, one may have somewhat generic Higgs effect in the 5D SLIV theory
which breaks the starting internal symmetry to its diagonal subgroups that
we discuss in detail in section 4. The most successful implementation of
this phenomena may appear in Grand Unified Theories considered ab initio in
the five-dimensional spacetime. As a result, these theories have to be
naturally broken down to the Standard Model at the 5D Lorentz violation
scale $\mathrm{M}$. And finally in section 5 we conclude.

\section{Emergent QED stemming from 5D spacetime}

For the reader's convenience, we will separate further discussion into the
particular steps that are needed for a final formulation of the emergent QED
theory in four dimensions.

\subsection{Five-dimensional QED with vector field constraints}

We start considering an Abelian $U(1)$ vector field theory in the 5D
Minkowski spacetime with an action

\begin{equation}
S=\int \mathrm{L}_{5D}d^{4}xdy  \label{ac}
\end{equation}%
where $x^{\mu }$ are conventional 4D coordinates and $y$ describes an extra
dimension (which we refer to as the fifth coordinate). The Lagrangian 
\textrm{L}$_{5D}$ is a conventional QED Lagrangian which according to our
philosophy also includes some covariant constraint put on five-dimensional
vector field $A_{\overline{\mu }}$. This may be implemented, as in the above
4D spacetime case (\ref{lag}), through an appropriate invariant Lagrange
multiplier term so that the Lagrangian \textrm{L}$_{5D}$ without matter
looks as 
\begin{equation}
\mathrm{L}_{5D}=-\frac{1}{4}F_{\overline{\mu }\overline{\nu }}F^{\overline{%
\mu }\overline{\nu }}-\frac{\mathrm{\lambda }}{2}\left( A_{\overline{\mu }%
}A^{\overline{\mu }}-\mathrm{n}^{2}\mathrm{M}_{5}^{2}\right) ,\text{ \ }%
\mathrm{n}^{2}=\mathrm{n}_{\overline{\mu }}\mathrm{n}^{\overline{\mu }}=\pm 1%
\text{ .}  \label{L5d}
\end{equation}%
where $\overline{\mu },\overline{\nu }$ are 5D indices, while $\mu ,\nu $
are 4D indices ($\overline{\mu },\overline{\nu }=\mu ,\nu ;5=0,1,2,3;5$).
The $\mathrm{\lambda }(x,y)$ is the Lagrange multiplier function, while the
mass parameter $\mathrm{M}_{5}$ stands for the mass scale where the 5D
Lorentz invariance is proposed to appear spontaneously broken along the
vacuum direction given now by a properly-oriented 5D unit vector $\mathrm{n}%
_{\overline{\mu }}$ which describes both of the 5D Lorentz violation cases
(timelike $\mathrm{n}^{2}=1$ or spacelike $\mathrm{n}^{2}=-1$) just by
analogy with the known 4D constraints (\ref{const}) discussed above. To see
more detail one has to come to conventional four dimensions. Some lessons
which can be retrieved from this tour may appear rather interesting for the
5D SLIV.

Assuming that the extra dimension is compactified as a circle of a radius of 
$R$, so that $y$ $\equiv $ $R\theta $, where $\theta $ is an angular
coordinate $-\pi \leq \theta \leq \pi $, we put the periodicity condition on
the starting 5D vector gauge fields taken, $A_{\overline{\mu }}(x,\theta
)=A_{\overline{\mu }}(x,\theta +2\pi )$. This allows for a Fourier expansion
as

\begin{equation}
A_{\overline{\mu }}(x,\theta )=\frac{1}{\sqrt{2\pi R}}A_{\overline{\mu }%
}^{0}(x)+\sum_{s=1}^{\infty }\frac{1}{\sqrt{\pi R}}\left[ A_{\overline{\mu }%
}^{s}(x)\cos (s\theta )+\widehat{A}_{\overline{\mu }}^{s}(x)\sin (s\theta )%
\right] \text{ }  \label{SW'}
\end{equation}%
where the first term in square bracket describes the modes being even under
reflection of the fifth coordinate ($\theta \rightarrow -$ $\theta $), while
the second one describes the modes which are odd under that reflection. Upon
putting $A_{\overline{\mu }}(x,\theta )$ into the action (\ref{ac}) and
integration over the extra dimension one gets for kinetic terms of the 4D
vector field components 
\begin{equation}
\mathrm{L}_{4D,kin}=\sum_{s=0}\left[ -\frac{1}{4}F_{\mu \nu }^{s}F^{s,\mu
\nu }+\frac{1}{2}\left( \partial _{\mu }A_{5}^{s}-\frac{s}{R}\widehat{A}%
_{\mu }^{s}\right) ^{2}\right] +(A\leftrightarrow \widehat{A})  \label{k}
\end{equation}%
where taking the fifth-coordinate derivative we have used, as prescribed
above, $\partial /\partial y=(1/R)\partial /\partial \theta $. One can see
that the terms within round brackets mix even and odd modes. These
combinations due to the starting gauge invariance of the 5D theory%
\begin{equation}
A_{\overline{\mu }}\rightarrow A_{\overline{\mu }}+\partial _{\overline{\mu }%
}\alpha (x,\theta )  \label{ga}
\end{equation}%
provides the mass term arrangement for KK towers. Indeed, with a general
parametrization (\ref{SW'})\ taken for gauge parameter $\alpha $ one has
from (\ref{ga}) for even KK modes 
\begin{equation}
A_{\mu }^{s}\rightarrow A_{\mu }^{s}+\partial _{\mu }\alpha ^{s}(x)\text{ ,
\ }A_{5}^{s}\rightarrow A_{5}^{s}-\frac{s}{R}\widehat{\alpha }^{s}(x)
\label{ga1}
\end{equation}%
and similarly for odd modes. Now, using this gauge freedom to diagonalize
the mixed terms in \textrm{L}$_{5D,kin}$ by proper fixing the gauges%
\begin{equation}
\alpha ^{s}=-(R/s)\widehat{A}_{5}^{s},\text{ }\widehat{\alpha }%
^{s}=-(R/s)A_{5}^{s}\text{\ }  \label{arr}
\end{equation}%
one finally gets 
\begin{equation}
\mathrm{L}_{4D,kin}=\sum_{s=0}\left[ -\frac{1}{4}F_{\mu \nu }^{s}F^{s,\mu
\nu }+\frac{1}{2}\left( \frac{s}{R}\right) ^{2}A_{\mu }^{s}A^{s,\mu }+\frac{1%
}{2}\left( \partial _{\mu }A_{5}^{0}\right) ^{2}\right] +(A\leftrightarrow 
\widehat{A}).  \label{4d}
\end{equation}%
Hence, the only massless vector field is given by the zero mode $A_{\mu }^{0}
$, while all KK modes acquire a mass by absorbing the scalars $A_{5}^{s}$.
This resembles the Higgs mechanism with $A_{5}^{s}$ playing the role of the
Goldstone bosons associated to the spontaneous 5D spacetime isometry
breaking \cite{36}. Remarkably, upon the gauge fixing arrangement (\ref{arr}%
) made for non-zero KK modes there still remains the $U(1)$ gauge symmetry
in the effective 4D theory with a massless gauge field $A_{\mu }^{0}$. Apart
from that, the massless scalar $A_{5}^{0}$ is also survived. However, this
extra degree of freedom appearing at zero level can be projected out from
the theory if the starting vector field component $A_{5}(x,\theta )$ is
chosen to be odd under the reflection $\theta \rightarrow -$ $\theta $
mentioned above.

\subsection{Looking for a natural compactification}

An adequate compactification certainly is a point of our special interest in
connection to the 5D SLIV. As is well-known, the necessity of producing
chiral fermions in four dimensions requires, on general grounds, to consider
the orbifold compactification $S_{1}/Z_{2}$ rather than a simple
compactification on a circle. This orbifold compactification consists in
fact of projecting a circular extra dimension onto a line with two fixed
points, $\theta =0$ and $\theta =\pi $ (or $y=0$ and $y=\pi R$ for the\ $y$
coordinate). It removes the unwanted fermionic degrees of freedom, allowing
for an existence of chiral fermions \cite{app, pil, hoop}. Actually, one has
to start with two 5D Dirac fermion fields to get independent left-handed and
righ-handed chiral modes in four dimensions: one field $\Psi _{1}$ which has
the \ quantum numbers of the left-handed spinor, and one field $\Psi _{2}$
with the quantum numbers of the right-handed spinor\footnote{%
Speaking about quantum numbers we have in mind the Standard Model extension
of our present QED framework.}. To exclude the additional degrees of freedom
one formulates the theory on an orbifold so that $\Psi _{1}$ is required to
be odd under the $\theta \rightarrow -$ $\theta $ orbifold symmetry, while $%
\Psi _{2}$ to be even 
\begin{equation}
\Psi _{1}(x,\theta )=-\gamma ^{5}\Psi _{1}(x,-\theta )\text{ , \ \ }\Psi
_{2}(x,\theta )=\gamma ^{5}\Psi _{2}(x,-\theta )  \label{orb}
\end{equation}%
\ \ In these fermion fields initially having a general form (\ref{SW'}) only
remain the parts 
\begin{eqnarray}
\Psi _{1,2}(x,\theta ) &=&\frac{1}{\sqrt{\pi R}}\Psi _{L1,R2}^{0}(x)  \notag
\\
&&+\sum_{s=1}\sqrt{\frac{2}{\pi R}}\left[ \Psi _{L1,R2}^{s}(x)\cos (s\theta
)+\widehat{\Psi }_{R1,L2}^{s}(x)\sin (s\theta )\right] ,  \label{or}
\end{eqnarray}%
respectively\footnote{%
The normalization of the fermion field here and all other fields everywhere
below is now chosen in accordance with an assumption that the range for the
angle variable $\theta $ to be from $0$ to $\pi $.}. As a result, their
higher KK modes are 4-dimensional vector-like fermions, while the zero modes
are chiral ones being properly determined by the chirality projectors $(1\mp
\gamma ^{5})/2$. As usual, their gauge couplings are in fact related
separately to each of these fermions, whereas in the Yukawa coupling they
"work" together\footnote{%
Note that generally after integrating over the fifth coordinate the sum over
KK number $s$ of the vector \ and matter fields in kinetic or interaction
terms in the\ corresponding effective 4D Lagrangian must be zero since this
is just conservation of the fifth dimension momentum. This conservation law,
being in essence the translational invariance along the extra dimension,
appears as an internal symmetry in the 4D KK decomposition, with internal
charges, $s$. Although the introduction of orbifold compactifications breaks
the above-mentioned symmetry related to conservation of the fifth dimension
momentum, a subgroup of the KK number conservation known as KK parity still
remains. In our 5D case compactified on an $S_{1}/Z_{2}$ orbifold, the KK
parity is the $Z_{2}$ symmetry and can be simply written as $P=(-1)^{s}$
where $s$ denotes the $s$-th KK mode. Thus, only modes with odd KK number
are charged.}.

At this point one must specify how all other fields transform under the
proposed orbifold projection. Specifically for vector fields, one usually
requires the "asymmetrical" compactification \cite{app, pil, hoop} according
to which the ordinary four components of the 5D vector field $A_{\overline{%
\mu }}(x,\theta )$ are even under the orbifold transformation, whereas its
fifth component is odd. This allows in the gauge invariant theory context to
completely remove this component from the theory excluding its zero mode $%
A_{5}^{0}$ by orbifold projection and gauging away the higher ones $A_{5}^{s}
$, as was discussed above. Thus, only massless ground modes $A_{\mu }^{0},$
as the Standard Model gauge field candidates, and massive vector KK towers $%
A_{\mu }^{s}$ ($s=1,2,...$) are left in the theory. However, as one can
readily see, such a procedure differently treating the vector field
components\ explicitly breaks the starting 5D Lorentz invariance that is
hardly acceptable if one tries to break it spontaneously. Thus, we propose,
in direct contrast to a common practice, the "symmetrical" compactification
in which all the 5D vector field components are even under the orbifold
transformation%
\begin{equation}
A_{\overline{\mu }}(x,-\theta )=A_{\overline{\mu }}(x,\theta ),\text{ }%
\overline{\mu }=0,1,2,3;5  \label{sym}
\end{equation}%
This, as we see below, may naturally conserve the 5D symmetrical form of all
possible non-derivative terms in the starting Lagrangian (\ref{L5d})
including the proposed vector field constraint terms that induce the SLIV.
Interestingly, such a "partially increased" Lorentz invariance significantly
reduces an effective gauge symmetry appearing for vector field components
after compactification. Indeed, for the vector field kinetic terms one has
now (when all the orbifold-asymmetrical vector field components vanish, $%
\widehat{A}_{\overline{\mu }}^{s}=0$) 
\begin{equation}
\mathrm{L}_{4D,kin}=\sum_{s=0}\left[ -\frac{1}{4}F_{\mu \nu }^{s}F^{s,\mu
\nu }+\frac{1}{2}(\partial _{\mu }A_{5}^{s})^{2}+\frac{1}{2}\left( \frac{s}{R%
}A_{\mu }^{s}\right) ^{2}\right] \text{ }  \label{k2}
\end{equation}%
and gauge symmetry (\ref{ga1}) for KK states, both massive vectors $A_{\mu
}^{s}$ ($s=1,2,...$) and massless scalars $A_{5}^{s}$ ($s=0,1,2,...$), does
not work any longer. Only standard gauge invariance for massless ground
vector modes $A_{\mu }^{0}$ holds%
\begin{equation}
A_{\mu }^{0}\rightarrow A_{\mu }^{0}+\partial _{\mu }\alpha ^{0}(x)\text{ }
\label{ga2}
\end{equation}%
which looks as if the 5D gauge function $\alpha $ in (\ref{ga}) would not
depend on the fifth coordinate and, therefore, only its ground component $%
\alpha ^{0}$ were nonzero. These states are completely decouple from each
other. Whereas zero vector field modes being protected by the above gauge
invariance are left massless, the massless scalars become eventually massive
through all the radiative corrections involved. Thus, they seem not to
produce serious difficulties for the model, as it could happen if the
extended gauge symmetry (\ref{ga1}) providing their masslessness remained.
Note also that, though the starting stress-tensor $F_{\overline{\mu }%
\overline{\nu }}$ does not look invariant under the "symmetrical" orbifold
transformation (\ref{sym}) the final Lagrangian \textrm{L}$_{4D,kin}$
appearing upon the compactification really does\footnote{%
Notably, though the derivative along the extra dimension is not invariant
under orbifold reflection ($\partial _{5}\rightarrow -\partial _{5}$), it is
actually replaced by $\partial _{5}$ $\rightarrow $ $-is/R$ in the Fourier
decomposition of the stress-tensor $F_{\overline{\mu }\overline{\nu }}$ and
then is modulo-squared in the kinetic terms so that the Lagrangian (\ref{k2}%
) appears perfectly invariant.}.

\subsection{Spacetime symmetry breaking phase}

Let us now turn to the Lagrange multiplier term in the starting Lagrangian (%
\ref{L5d}) which is proposed to cause spontaneous 5D Lorentz violation.
Taking also the multiplier function to be, as all other fields but fermions,
symmetrical under orbifold transformation, 
\begin{equation}
\mathrm{\lambda }(x,\theta )=\frac{1}{\sqrt{\pi R}}\left[ \mathrm{\lambda }%
^{0}(x)+\sqrt{2}\sum_{s=1}\mathrm{\lambda }^{s}(x)\cos (s\theta )\right]
\end{equation}%
and varying the action with respect to all KK components, $\mathrm{\lambda }%
^{0}$ and $\mathrm{\lambda }^{s}$, one has after integration over the angle $%
\theta $

\begin{eqnarray}
A_{\overline{\mu }}^{0}A^{0\overline{\mu }}+\sum_{s=1}A_{\overline{\mu }%
}^{s}A^{s\overline{\mu }} &=&\mathrm{n}^{2}\mathrm{M}^{2}\text{ ,}  \notag \\
\sqrt{2}A_{\overline{\mu }}^{0}A^{s\overline{\mu }}+\sum_{s^{\prime }=1}A_{%
\overline{\mu }}^{s-s^{\prime }}A^{s^{\prime }\overline{\mu }} &=&0\text{ \ (%
}s=1,2,...\text{), }  \label{constr}
\end{eqnarray}%
respectively\footnote{%
Note that, for convenience and to emphasize the KK mode number conservation$%
^{5}$, we formally included in the sums here and everywhere below the 4D KK\
modes $A_{\mu }^{S}$ with possible negative numbers $S$ as well, though for
the "symmetrical" orbifold compactification taken one has $A_{\mu
}^{S}=A_{\mu }^{\left\vert S\right\vert }$ for every value of $S$.}. The
evident relation was also used between 4D and 5D mass scales, $\mathrm{M}%
^{2}=(\pi R)\mathrm{M}_{5}^{2}$. The first constraint in (\ref{constr})
resembles the 4D constraint discussed above (\ref{const}), while the others
are new. Actually we have one constraint for each vector field mode, $A_{%
\overline{\mu }}^{0}$ and $A_{\overline{\mu }}^{s}$ ($s=1,2,...$). One can
see that, though the "symmetrical" orbifold compactification (\ref{sym})
taken above for vector field breaks the 5D Lorentz invariance in the
Lagrangian (\ref{k2}), it perfectly conserves the 5D invariant form of the
constraints (\ref{constr}). They lead in turn to spontaneous violation of 5D
Lorentz symmetry and\ production of massless 4D vector bosons as the
corresponding Goldstone modes which, due to the lesser symmetry of the total
Lagrangian, are in essence the pseudo-Goldstone modes (see below). In
contrast, a conventional "asymmetrical" orbifold compactification for
starting 5D vector field $A_{\overline{\mu }}(x,\theta )$ would explicitly
break the Lagrangian 5D form invariance of these constraints\footnote{%
In this case the first terms in the constraints (\ref{constr}) containing
zero modes would have only 4D invariant form, being just $(A_{\mu }^{0})^{2}$
and $\sqrt{2}A_{\mu }^{0}A_{\mu }^{s}$, respectively.} and make such an
implementation impossible.

Applying the same constraints, as they are given in (\ref{constr}), to a
possible VEV (vacuum expectation value) of the 5D vector field $A_{\overline{%
\mu }}(x,\theta )$ expanded in a Fourier cosine series in (\ref{SW'})\ one
could conclude that this VEV may only develop on its ground mode rather than
the higher KK ones in order not to be dependent on the extra dimension
coordinate. Thus, the starting 5D Lorentz symmetry will break due to the VEV
developed solely on the zero modes $\mathbf{A}_{\overline{\mu }}^{0}$. As to
the particular spacetime component $\overline{\mu }$ on which this VEV may
develop, we propose that just the space-like 5D SLIV case ($\mathrm{n}%
^{2}=-1 $) is realized in the present model. Particularly, this symmetry
will indeed be spontaneously broken to ordinary Lorentz invariance%
\begin{equation}
SO(1,4)\rightarrow SO(1,3)  \label{brr}
\end{equation}%
at a scale $\mathrm{M}$ 
\begin{equation}
\left\langle A_{\overline{\mu }}\right\rangle \text{ }=\mathrm{n}_{\overline{%
\mu }}\mathrm{M}\text{, \ }\mathrm{n}^{2}=-1
\end{equation}%
with the vacuum direction given now by the `unit' vector $\mathrm{n}_{%
\overline{\mu }}$ with the only non-zero component $\mathrm{n}_{\overline{%
\mu }}=g_{\overline{\mu }5}$ just along the extra dimension. One can write
again, as in the 4D case mentioned above (\ref{gol}), the ground vector
field expansion around vacuum configuration stemming from the upper
constraint in (\ref{constr}) 
\begin{equation}
A_{\overline{\mu }}^{0}=a_{\overline{\mu }}+\mathrm{n}_{\overline{\mu }}%
\sqrt{\mathrm{M}^{2}+a^{2}+(A^{s})^{2}},\text{ \ \ }  \label{e}
\end{equation}%
where summation over all repeated indices is taken%
\begin{equation}
a^{2}\equiv a_{\overline{\mu }}a^{\overline{\mu }},\text{ }(A^{s})^{2}\equiv
\sum_{s=1}A_{\overline{\mu }}^{s}A^{\overline{\mu }s}\text{ ,}
\end{equation}%
and also the orthogonality condition for the emergent pseudo-Goldstone modes 
$a_{\overline{\mu }}$ 
\begin{equation}
\mathrm{n}^{\overline{\mu }}a_{\overline{\mu }}=0
\end{equation}%
is supposed. Meanwhile, the effective Higgs field\ in the model is given by%
\begin{equation}
\mathrm{H}=\mathrm{n}^{\overline{\mu }}A_{\overline{\mu }}^{0}=A_{5}^{0}=%
\sqrt{\mathrm{M}^{2}+a^{2}+(A^{s})^{2}}  \label{h}
\end{equation}%
Note, as mentioned above, that, while the constraints (\ref{constr}) are
formally 5D Lorentz invariant, the vector field kinetic terms in the 4D
Lagrangian (\ref{k2}) and also all interaction terms involved possesses only
ordinary 4D Lorentz invariance once the compactification occurs. This means
that all the 4D modes $a_{\mu }$ appeared in the above expansion (\ref{e})
are in fact pseudo-Goldstone bosons (PGB) related to the accidental symmetry
breaking (\ref{brr}) of the constraints (\ref{constr}), rather than true
Goldstone vector modes. Remarkably, in contrast to the familiar scalar PGB
case \cite{GLA}, these vector PGBs remain strictly massless being protected
by gauge invariance (\ref{ga2}) surviving after our "symmetrical" orbifold
compactification for massless ground vector mode $A_{\mu }^{0}$ which
coincides with the PGB state $a_{\mu }$ in the expansion (\ref{e}).

\subsection{Emergent QED: some immediate consequences}

Finally, one can see \ that the QED theory emerging from the 5D spacetime
has a quite simple form, though contains some extra interaction terms.
Indeed, separating ground modes and heavy KK modes in the 4D Lagrangian (\ref%
{k2}) and putting the expansion (\ref{e}) one eventually comes to the
emergent QED theory in four dimensions (matter terms are omitted) 
\begin{eqnarray}
\mathrm{L}_{em}(a,\text{\textrm{\ }}A^{s}) &=&-\frac{1}{4}f_{\mu \nu }f^{\mu
\nu }+\frac{1}{2}(\partial _{\mu }\mathrm{H})^{2}  \notag \\
&&+\sum_{s=1}\left[ -\frac{1}{4}F_{\mu \nu }^{s}F^{s,\mu \nu }+\frac{1}{2}%
(\partial _{\mu }A_{5}^{s})^{2}+\frac{1}{2}\left( \frac{s}{R}A_{\mu
}^{s}\right) ^{2}\right]  \label{mr}
\end{eqnarray}%
where we have introduced stress-tensor for PGB modes, $f_{\mu \nu }=\partial
_{\mu }a_{\nu }-\partial _{\nu }a_{\mu }$, and "kinetic" term for effective
Higgs field $\mathrm{H}$ (\ref{h}). The latter, when is properly expanded,
gives all possible multi-boson couplings 
\begin{eqnarray}
\frac{1}{2}(\partial _{\mu }\mathrm{H})^{2} &=&\frac{1}{2}\frac{\left(
a_{\rho }\partial _{\mu }a^{\rho }+\sum_{s=1}A_{\overline{\rho }%
}^{s}\partial _{\mu }A^{s,\overline{\rho }}\right) ^{2}}{\mathrm{M}%
^{2}+a^{2}+(A^{s})^{2}}  \notag \\
&=&\frac{1}{2\mathrm{M}^{2}}\left( a_{\rho }\partial _{\mu }a^{\rho
}+\sum_{s=1}A_{\overline{\rho }}^{s}\partial _{\mu }A^{s,\overline{\rho }%
}\right) ^{2}\left[ 1+\sum_{n=1}^{\infty }\left( -\frac{a^{2}+(A^{s})^{2}}{%
\mathrm{M}^{2}}\right) ^{n}\right]  \label{mm}
\end{eqnarray}%
in addition to conventional QED interactions. Thus, starting from the order
of $\mathcal{O}(1/\mathrm{M}^{2})$ there appear some direct photon-photon
scattering couplings and also couplings photons with heavy KK modes in the
emergent QED which, therefore, possess only a partial gauge invariance. In
contrast to the known 4D Nambu model \cite{nambu}, where the direct
photon-photon scattering amplitudes are always cancelled by accompanying
longitudinal photon exchange terms, in the 5D model they appear alone and
consequently are survived. Therefore, their observation could shed light on
the emergent nature of QED stemming from 5D spacetime. Interestingly, due to
the orbifold symmetry taken for vector fields and fermions (\ref{sym}, \ref%
{orb}) the matter fields (both fermions and scalars) when being introduced
into the 5D QED do not produce any new "emergent" couplings for their ground
modes. So, only vector fields, photons and heavy KK modes, acquire some
extra direct multi-boson interactions (\ref{mm}) when the effective Higgs
field $\mathrm{H}$ related to the 5D Lorentz violation is properly expanded
in the basic Lagrangian (\ref{mr}).

Another crucial prediction of the emergent QED is an existence of the
unabsorbed fifth-direction non-zero modes $A_{5}^{s}$ ($s=1,2,...$) being
massless at the tree-level. Due to KK parity$^{5}$ they can be only produced
by pairs from an ordinary matter being properly suppressed by the 5D Lorentz
violation scale $\mathrm{M}$ (as in the photon-photon scattering\ processes
given above in\ (\ref{mm})) or by the compactification mass $M_{KK}\sim 1/R$
(when such a process is caused by the heavy KK mode exchange). On the other
hand, any heavy KK state will now rapidly decay into the $A_{5}^{s}$ mode
plus ordinary matter that seems to invalidate the dark matter scenario
related to extra dimension \cite{hoop}. However, these $A_{5}^{s}$ modes
being no more protected by gauge invariance could in principle acquire large
masses through radiative corrections so that the lightest KK state may
appear rather stable to provide the measured dark matter density.

\section{Emergent Yang-Mills theory}

We now consider Yang-Mills theory in the five-dimensional Minkowski
spacetime with the vector field Lagrangian 
\begin{eqnarray}
\mathcal{L}_{5D} &=&-\frac{1}{4}Tr\left\vert \mathbf{F}_{\overline{\mu }%
\overline{\nu }}\right\vert ^{2}+\mathcal{\lambda }\left[ Tr(\mathbf{A}_{%
\overline{\mu }}\mathbf{A}^{\overline{\mu }})-\mathbf{n}^{2}\mathrm{M}%
_{5}^{2}\right] ,  \notag \\
\mathbf{F}_{\overline{\mu }\overline{\nu }} &=&\partial _{\overline{\mu }}%
\mathbf{A}_{\overline{\nu }}-\partial _{\overline{\nu }}\mathbf{A}_{%
\overline{\mu }}+ig_{5}[\mathbf{A}_{\overline{\mu }},\mathbf{A}_{\overline{%
\nu }}]\text{ \ }(\overline{\mu },\overline{\nu }=\mu ,\nu ;5)\text{ .}
\label{nal}
\end{eqnarray}%
This non-Abelian internal symmetry case is supposed to be given by a general
local group $G$ with generators $t^{i}$ ($[t^{i},t^{j}]=if^{ijk}t^{k}$ and $%
Tr(t^{i}t^{j})=\delta ^{ij}$ where $f^{ijk}$ are structure constants and $%
i,j,k=0,1,...,N-1$). The corresponding 5D vector fields\ which transform
according to its adjoint representation are given in the proper matrix form $%
\mathbf{A}_{\overline{\mu }}=\mathbf{A}_{\overline{\mu }}^{i}t^{i}$, while
the possible matter fields (fermions, for definiteness) could be presented
in the fundamental representation column $\mathfrak{\psi }^{r}$ ($%
r=0,1,...,d-1$) of $G$. \ According to our philosophy, the starting theory,
as in the above Abelian case, also contains some covariant constraint put on
5D vector field $\mathbf{A}_{\overline{\mu }}$ that causes at the scale $%
\mathrm{M}_{5}$ a spontaneous violation of the 5D Lorentz invariance
involved. This is arranged through the Lagrange multiplier term in the
Lagrangian(\ref{nal}) with the multiplier function $\mathfrak{\lambda }(x,y)$
depending in general on all five coordinates. The vacuum direction is given
now by a properly-oriented `unit' rectangular matrix $\mathbf{n}_{\overline{%
\mu }}^{i}$ which describes in general both of the 5D Lorentz violation
cases (timelike or spacelike)%
\begin{equation}
\mathbf{n}_{\overline{\mu }}=\mathbf{n}_{\overline{\mu }}^{i}t^{i}\text{, \ }%
\mathbf{n}^{2}=\mathbf{n}_{\overline{\mu }}\mathbf{n}^{\overline{\mu }}=\pm 1
\label{const1}
\end{equation}

Decomposing all fields in the Lagrangian (\ref{nal}) in a Fourier cosine
series along the fifth coordinate one has 
\begin{eqnarray}
\mathbf{A}_{\overline{\mu }}(x,\theta ) &=&\frac{1}{\sqrt{\pi R}}\left[ 
\mathbf{A}_{\overline{\mu }}^{0}(x)+\sqrt{2}\sum_{s=1}^{\infty }\mathbf{A}_{%
\overline{\mu }}^{s}(x)\cos (s\theta )\right] \text{ ,}  \notag \\
\mathfrak{\lambda }(x,\theta ) &=&\frac{1}{\sqrt{\pi R}}\left[ \mathcal{%
\lambda }^{0}(x)+\sqrt{2}\sum_{s=1}^{\infty }\mathcal{\lambda }^{s}(x)\cos
(s\theta )\right] \text{ ,}  \label{lm}
\end{eqnarray}%
where it was again proposed that they all five vector field components $%
\mathbf{A}_{\overline{\mu }}$, as well as the multiplier function $\mathfrak{%
\lambda }$, are even under orbifold transformation%
\begin{equation}
\mathbf{A}_{\overline{\mu }}(x,-\theta )=\mathbf{A}_{\overline{\mu }%
}(x,\theta )\text{, \ }\mathfrak{\lambda }(x,-\theta )=\mathfrak{\lambda }%
(x,\theta )\text{ .}  \label{sym1}
\end{equation}%
Then integrating the action over the angle $\theta $ and varying the
resulting 4D Lagrangian $\mathcal{L}_{4D}$ with respect to the zero and
higher KK modes, $\mathcal{\lambda }^{0}$ and $\mathcal{\lambda }^{s}$ of $\ 
$the multiplier function$\ \mathcal{\lambda }(x,\theta )$ one obtains all
possible constraints put on the properly normalized vector field 4D modes$%
^{7}$%
\begin{eqnarray}
Tr\left( \mathbf{A}_{\overline{\mu }}^{0}\mathbf{A}^{0\overline{\mu }%
}\right) +\sum_{s=1}Tr(\mathbf{A}_{\overline{\mu }}^{s}\mathbf{A}^{s%
\overline{\mu }}) &=&\mathbf{n}^{2}\mathrm{M}^{2}\text{ ,}  \notag \\
\sqrt{2}Tr\left( \mathbf{A}_{\overline{\mu }}^{0}\mathbf{A}^{s\overline{\mu }%
}\right) +\sum_{s^{\prime }=1}Tr\left( \mathbf{A}_{\overline{\mu }%
}^{s-s^{\prime }}\mathbf{A}^{s^{\prime }\overline{\mu }}\right)  &=&0\text{
\ (}s=1,2,...\text{). }  \label{CONST}
\end{eqnarray}%
where the evident relation was also used between 4D and 5D mass scales, $%
\mathrm{M}^{2}=(\pi R)\mathrm{M}_{5}^{2}$. Eventually, we have one
constraint for each vector field mode, $\mathbf{A}_{\overline{\mu }}^{0}$
and $\mathbf{A}_{\overline{\mu }}^{s}$ ($s=1,2,...$), while the final 4D
Lagrangian (with the Lagrange multiplier term omitted) may be written as%
\begin{eqnarray}
\mathcal{L}_{4D} &=&-\frac{1}{4}\sum_{s=0}Tr\left( \left\vert \overline{%
\mathbf{F}}_{\mu \upsilon }^{s}\right\vert ^{2}-2\left\vert \overline{%
\mathbf{F}}_{\mu 5}^{s}\right\vert ^{2}\right)   \notag \\
&&+\sum_{s=1}\mathcal{O}[(\mathbf{A}^{0})^{2}(\mathbf{A}^{s})^{2},(\mathbf{A}%
^{0}\mathbf{A}^{s})(\mathbf{A}^{s})^{2},(\mathbf{A}^{s})^{2}(\mathbf{A}%
^{s})^{2}]\text{ .}  \label{lag11}
\end{eqnarray}%
where we truncated the vector field covariant derivatives ignoring in the
stress-tensors $\overline{\mathbf{F}}_{\mu \upsilon }^{s}$ and $\overline{%
\mathbf{F}}_{\mu 5}^{s}$ the commutator terms for non-zero KK modes\
(appearing in the Lagrangian in the indicated orders) that is unessential
for the further analysis. One can see that, though the "symmetrical"
orbifold compactification (\ref{sym1}) taken above for vector field
multiplet $\mathbf{A}_{\overline{\mu }}$ breaks the 5D Lorentz invariance in
the Lagrangian (\ref{lag11}), it perfectly conserves the 5D invariant form
of the constraints (\ref{CONST}). They lead in turn to spontaneous violation
of 5D Lorentz symmetry and\ production of massless 4D vector bosons as the
corresponding pseudo-Goldstone modes related to the total symmetry breaking.

Let us consider this 5D SLIV phenomenon in more detail. Applying the same
constraints (\ref{CONST}) to possible VEV of \ the 5D vector field multiplet 
$\mathbf{A}_{\overline{\mu }}(x,\theta )$ expanded in a Fourier series in (%
\ref{lm})\ one could conclude that, as in the above Abelian case, this VEV
may only develop on its ground mode rather than the higher KK ones in order
not to be dependent on the extra dimension coordinate. Thus, the starting 5D
Lorentz symmetry breaks due to the VEV developed solely on the zero modes $%
\mathbf{A}_{\overline{\mu }}^{0}$. As to the particular spacetime component $%
\overline{\mu }$ on which this VEV may develop, we propose in what follows
the space-like 5D SLIV ($\mathbf{n}^{2}=-1$) in the theory, thus taking the
case $\mathbf{n}^{2}=-1$ in (\ref{const1}). However, there is one special
point in the non-Abelian theory framework (with an internal symmetry group $%
G $ introduced) that has been studied before in a conventional 4D spacetime 
\cite{jej, cj, wiki}. Namely, although we only propose the $SO(1,4)\times G$
invariance of the Lagrangian (\ref{lag11}), the vector field constraint (\ref%
{CONST}) (or, equally, some possible polynomial potential terms which could
be included into the starting Lagrangian $\mathcal{L}_{5D}$) possesses in
fact much higher accidental global symmetry $SO(N,4N)$ determined by the
dimensionality $N$ of the $G$ group adjoint representation to which the
vector field multiplet $\mathbf{A}_{\overline{\mu }}$ belong. This symmetry
is indeed spontaneously broken%
\begin{equation}
SO(N,4N)\rightarrow SO(N,4N-1)  \label{so}
\end{equation}%
at a scale $\mathrm{M}$ 
\begin{equation}
\left\langle \mathbf{A}_{\overline{\mu }}^{i}\right\rangle \text{ }=\mathbf{n%
}_{\overline{\mu }}^{i}\mathrm{M}  \label{vev}
\end{equation}%
with the vacuum direction given by the matrix $\mathbf{n}_{\mu }^{i}$
describing now the 5D space-like SLIV case, $\mathbf{n}^{2}=-1$. Without
loss of generality, this matrix can be written in the factorized
"two-vector" form $\mathbf{n}_{\overline{\mu }}^{i}=n_{\overline{\mu }}%
\mathbf{I}^{i}$ where $n_{\overline{\mu }}$ is the unit Lorentz vector which
is oriented in 5D spacetime so as to be parallel to the vacuum matrix $%
\mathbf{n}_{\overline{\mu }}^{i}$, while $\mathbf{I}^{i}$ is the unit vector
in the internal space ($\mathbf{I}^{i}\mathbf{I}^{i}=1$). This matrix $%
\mathbf{n}_{\mu }^{i}$ has in fact only one non-zero element subject to the
appropriate $SO(N,4N)$ rotation. This is, specifically, 
\begin{equation}
\mathbf{n}_{\overline{\mu }}^{i}=n_{\overline{\mu }}\mathbf{I}^{i}=g_{%
\overline{\mu }5}\delta ^{ii_{0}}\text{ ,}  \label{v}
\end{equation}%
provided that the vacuum expectation value (\ref{vev}) is developed along
the $i_{0}$ direction in the internal space and along the $\overline{\mu }=5$
direction, respectively, in the 5D Minkowski spacetime.

One can readily see that, in response to this breaking (\ref{so}) the $5N-1$
massless modes according to a number of broken generators are therefore
produced. Actually, due to the symmetry reduction in the
post-compactification Lagrangian (\ref{lag11}) all these Goldstone modes $%
\mathbf{a}_{\overline{\mu }}^{i}$ are in fact pseudo-Goldstone bosons
related to breaking of the accidental $SO(N,4N)$ symmetry of the SLIV
constraints (\ref{CONST}). They are excited along the directions being
orthogonal to the vacuum determined by the above unit vector $\mathbf{n}_{%
\overline{\mu }}^{i}$ 
\begin{equation}
\mathbf{n}_{\overline{\mu }}^{i}\mathbf{a}_{\overline{\mu }}^{i}=\mathbf{0}%
\text{ \ }(i=i_{0},1,...,N-1)  \label{oor}
\end{equation}%
These PGBs include $N$ four-component vector modes $\mathbf{a}_{\mu }^{i}$
which complete the adjoint vector field multiplet of the internal symmetry
group $G$. Again as in the Abelian case, these vector PGBs, in sharp
contrast to the familiar scalar PGB case \cite{GLA}, remain strictly
massless being protected by the non-Abelian gauge invariance in the final
Lagrangian (\ref{lag11}) where an actual symmetry $SO(1,3)\otimes G$ still
remains\footnote{%
Actually, the internal symmetry group $G$ eventually appears spontaneously
broken to its diagonal subgroups (see below).}. Apart from them, there are
the $N-1$ massless scalar modes $\mathbf{\phi }^{i}\equiv \mathbf{a}_{5}^{i}$
($i=1,...,N-1$). In contrast to vector bosons, they are not protected by any
gauge symmetry and consequently will get masses through the radiative
corrections.

One can write again, as in the above Abelian case (\ref{e}), the ground
vector field expansion around vacuum configuration stemming from the upper
constraint in (\ref{CONST}) 
\begin{equation}
\mathbf{A}_{\overline{\mu }}^{0i}=\mathbf{a}_{\overline{\mu }}^{i}+\mathbf{n}%
_{\overline{\mu }}^{i}\sqrt{\mathrm{M}^{2}-\mathbf{n}^{2}\mathbf{a}^{2}-%
\mathbf{n}^{2}(\mathbf{A}^{s})^{2}},\text{ \ \ }(\mathbf{a}^{2}\equiv 
\mathbf{a}_{\overline{\mu }}^{i}\mathbf{a}^{\overline{\mu }i},\text{ }(%
\mathbf{A}^{s})^{2}\equiv \mathbf{A}_{\overline{\mu }}^{si}\mathbf{A}^{%
\overline{\mu }si}\text{ })  \label{exp}
\end{equation}%
where summation over all repeated indices is taken, and also the
orthogonality condition (\ref{oor})\ for the emergent pseudo-Goldstone $%
\mathbf{a}_{\overline{\mu }}^{i}$ is supposed. Meanwhile, the effective
Higgs term in the expansion (\ref{exp})%
\begin{equation}
\mathcal{H}=\mathbf{A}_{\overline{\mu }}^{0i}\mathbf{n}_{\overline{\mu }%
}^{i}=\sqrt{\mathrm{M}^{2}-\mathbf{n}^{2}\mathbf{a}^{2}-\mathbf{n}^{2}(%
\mathbf{A}^{s})^{2}}=\mathrm{M}+\mathcal{O}(\mathbf{a}^{2}/\mathrm{M},\text{ 
}(\mathbf{A}^{s})^{2}/\mathrm{M})  \label{deco}
\end{equation}%
induces masses for some set of vector fields inside of the multiplet $%
\mathbf{a}_{\mu }^{i}$. Putting the expansion (\ref{exp}) into the
Lagrangian (\ref{lag11}) one eventually comes to the emergent 4D Yang-Mills
theory stemming from the 5D spacetime with all possible vector field
couplings involved. They are given for the ground modes by the truncated
stress-tensors presented in the Lagrangian (\ref{lag11}) 
\begin{equation}
\overline{\mathbf{F}}_{\mu \upsilon }^{0}=\partial _{\mu }\mathbf{a}%
_{\upsilon }-\partial _{\upsilon }\mathbf{a}_{\mu }+ig[\mathbf{a}_{\mu },%
\mathbf{a}_{\upsilon }]  \label{oo}
\end{equation}%
and \ 
\begin{equation}
\overline{\mathbf{F}}_{\mu 5}^{0}=\partial _{\mu }(\mathbf{\phi +l}\mathcal{H%
})+ig\left( [\mathbf{a}_{\mu },\mathbf{\phi }]+\mathcal{H}[\mathbf{a}_{\mu },%
\mathbf{l}]\right)  \label{55}
\end{equation}%
{\Huge \ }while for the higher modes ($s=1,2,...$) by the truncated tensors 
\begin{equation}
\frac{\overline{\mathbf{F}}_{\mu \upsilon }^{s}}{\sqrt{2}}=\partial _{\mu }%
\mathbf{A}_{\nu }^{s}-\partial _{\upsilon }\mathbf{A}_{\mu }^{s}+ig\left( [%
\mathbf{A}_{\mu }^{s},\mathbf{a}_{\upsilon }]+[\mathbf{a}_{\mu },\mathbf{A}%
_{\upsilon }^{s}]\right)  \label{ooo}
\end{equation}%
and 
\begin{equation}
\frac{\overline{\mathbf{F}}_{\mu 5}^{s}}{\sqrt{2}}=\partial _{\mu }\mathbf{A}%
_{5}^{s}+ig\left( [\mathbf{A}_{\mu }^{s},\mathbf{\phi }]+\mathcal{H}[\mathbf{%
A}_{\mu }^{s},\mathbf{l}]\right) -i\frac{s}{R}\mathbf{A}_{\mu }^{s}\text{.}
\label{555}
\end{equation}%
where an effective 4D gauge coupling constant $g=g_{5}/\sqrt{\pi R}$ has
been introduced. Such a form of these tensors\ readily follows upon an
integration of the corresponding action over the extra dimension where also
the normalization for ground and higher modes is properly taken into account.

Note that the starting theory (\ref{nal}) without the Lagrange multiplier
term is invariant under the 5D non-Abelian gauge transformations of the
vector field multiplet 
\begin{equation}
\mathbf{A}_{\overline{\mu }}^{\prime }=\mathbf{A}_{\overline{\mu }}+i[%
\mathbf{\alpha },\mathbf{A}_{\overline{\mu }}]+\partial _{\overline{\mu }}%
\mathbf{\alpha }  \label{gt0}
\end{equation}%
where the gauge parameter $\mathbf{\alpha =\alpha }^{i}t^{i}$ is also
proposed to have a symmetrical cosine expansion as the vector field
components $\mathbf{A}_{\overline{\mu }}$. After compactification for a
particular $s$\ component ($s=0,1,...$) it turns to%
\begin{eqnarray}
\delta \mathbf{A}_{\mu }^{s} &=&i\sum_{s^{\prime }=0}[\mathbf{\alpha }%
^{s^{\prime }},\mathbf{A}_{\mu }^{s-s^{\prime }}]+\partial _{\mu }\mathbf{%
\alpha }^{s},  \notag \\
\delta \mathbf{A}_{5}^{s} &=&i\sum_{s^{\prime }=0}[\mathbf{\alpha }%
^{s^{\prime }},\mathbf{A}_{5}^{s-s^{\prime }}]\text{ }  \label{gt1}
\end{eqnarray}%
showing that in the "rotation" part of each KK mode (ground state or higher
KK mode) contribute all other states as well. Remarkably, for the
symmetrical orbifold compactification taken the fifth-direction modes $%
\mathbf{A}_{5}^{s,i}$ are only rotated under the internal symmetry group
transformations thus behaving themselves as the matter fields rather than
the gauge field components\footnote{%
This will allow us later (section 4) to treat its ground mode multiplet $%
\mathbf{A}_{5}^{0,i}$ ($i=0,1,...,N-1$) as an independent adjoint Higgs
field multiplet in the theory considered.}. This means that they can not be
gauged away from the theory. Therefore, as in the Abelian case, one has
eventually, apart from the massless ground modes $\mathbf{A}_{\overline{\mu }%
}^{0,i}=\mathbf{a}_{\overline{\mu }}^{i}=(\mathbf{a}_{\mu }^{i},\mathbf{\phi 
}^{i})$, the massive vector KK modes $\mathbf{A}_{\mu }^{s,i}$ and the
massless "scalars" $\mathbf{A}_{5}^{s,i}$ ($s=1,2,...$). Moreover, the
latter modes $\mathbf{A}_{\mu }^{s,i}$ and $\mathbf{A}_{5}^{s,i}$ break in
essence the starting gauge invariance (\ref{gt1}) down to a conventional
gauge invariance related solely to the ground vector field modes in the 4D
Lagrangian (\ref{lag11}) 
\begin{equation}
\delta \mathbf{A}_{\mu }^{s}=i[\mathbf{\alpha }^{0},\mathbf{A}_{\mu
}^{s}]+\partial _{\mu }\mathbf{\alpha }^{s}\delta ^{s0}\text{\ , \ \ \ }%
\delta \mathbf{A}_{5}^{s}=i[\mathbf{\alpha }^{0},\mathbf{A}_{5}^{s}]\text{ }
\label{gt2}
\end{equation}%
while all other modes are only "rotated" by the internal group symmetry
generators. Actually, again as in the above Abelian case, this looks as if
the 5D gauge function $\mathbf{\alpha }$ in (\ref{gt0}) would not depend on
the fifth coordinate and, therefore, only its ground component $\mathbf{%
\alpha }^{0}$ were nonzero. This restricted gauge invariance makes it
possible to uncover some direct observational effects related to the 5D
SLIV, in contrast to the completely hidden spontaneous Lorentz violation
appearing in a conventional 4D spacetime \cite{jej, cj, wiki}.

These effects are essentially determined by the stress-tensors $\overline{%
\mathbf{F}}_{\mu 5}^{0}$ and $\overline{\mathbf{F}}_{\mu 5}^{s}$ (\ref{55}, %
\ref{555}) related to the fifth direction. Again, as in Abelian case, apart
from conventional gauge couplings presented in (\ref{oo}), one has direct
multi-boson (multi-photon in particular) couplings following from the
"kinetic" term of the effective scalar $\mathcal{H}$ in the Lagrangian (\ref%
{lag11})%
\begin{equation}
\frac{1}{2}(\partial _{\mu }\mathcal{H})^{2}=\frac{1}{2}\frac{\left( \mathbf{%
a}_{\rho }^{i}\partial _{\mu }\mathbf{a}^{\rho i}+\sum_{s=1}\mathbf{A}_{%
\overline{\rho }i}^{s}\partial _{\mu }\mathbf{A}^{s,\overline{\rho }%
i}\right) ^{2}}{\mathrm{M}^{2}+\mathbf{a}^{2}+(\mathbf{A}^{s})^{2}}\text{ .}
\label{hh}
\end{equation}%
As a matter of fact, there appear an infinite number of the properly
suppressed direct vector boson-boson scattering couplings and also couplings
of these ground mode bosons with heavy KK towers in the Lagrangian. Thus,
the emergent Yang-Mills theory, just as the emergent QED considered above,
actually possess only a partial gauge invariance whose observation could be
of primary interest. Likewise, an existence of the unabsorbed
fifth-direction non-zero modes $\mathbf{A}_{5}^{s,i}$ ($s=1,2,...$) being
massless at the tree-level appears as somewhat unavoidable prediction of the
model. Again, due to KK parity they will be only produced by pairs from an
ordinary matter being properly suppressed by the 5D Lorentz violation scale $%
\mathrm{M}$, as is shown in (\ref{hh}) for their possible production in
boson-boson scattering and other processes. Also, any heavy KK state will
now rapidly decay into the $\mathbf{A}_{5}^{s,i}$ modes plus ordinary
matter. However, all these massless fifth-direction states in the model, the
ground modes $\mathbf{A}_{5}^{0,i}=\mathbf{a}_{5}^{i}$ and towers $\mathbf{A}%
_{5}^{s,i}$, being no more protected by the restricted gauge invariance (\ref%
{gt2}) could in principle acquire quite large masses (see some details in
section 4), thus escaping the direct observation.

All these predictions may be equally expected from both Abelian and
non-Abelian theory cases. However, there is one point being particularly
specific to the emergent Yang-Mills theory. This is the generic Higgs
mechanism appearing in a non-Abelian theory which leads to an automatic
internal symmetry reduction when the 5D symmetry spacetime symmetry
spontaneously breaks down to a conventional Lorentz invariance.

\section{Internal symmetry reduction in four dimensions}

We have seen above that if the starting 5D theory possesses some non-Abelian
internal symmetry $G$ this symmetry, simultaneously with the underlying
spacetime symmetry, occurs spontaneously broken. As a result, some
pseudo-Goldstone vector bosons emerging during symmetry breaking process (%
\ref{so}) may acquire masses. This breaking itself appears similar to the
breaking which is usually induced by an introduction into the theory of an
appropriate adjoint scalar field multiplet. Now such a multiplet origins
from extra vector field components. Particularly, in the 5D theory the role
of such a scalar field multiplet plays the multiplet composed from the fifth
component $\left( \mathbf{A}_{5}^{0}\right) ^{i}$ of the zero-mode vector
field $\left( \mathbf{A}_{\overline{\mu }}^{0}\right) ^{i}$, whose VEV is
given by the equations (\ref{vev}, \ref{v}) depending on the direction $%
i=i_{0}$ in the internal space along which $G$ symmetry appears broken. They
through their covariant derivatives give masses to the 4D ground vector
field modes $\mathbf{a}_{\mu }^{i}$. Therefore, one may have the Higgs
effect in the 5D SLIV theory without a specially introduced Higgs field.

Let us consider it in more detail. Rewriting the starting field expansion (%
\ref{exp}) for particular components we receive 
\begin{eqnarray}
\left( \mathbf{A}_{\mu }^{0}\right) ^{i} &=&\mathbf{a}_{\mu }^{i},  \notag \\
\left( \mathbf{A}_{5}^{0}\right) ^{i} &=&\mathbf{\phi }^{i}+\mathbf{l}^{i}%
\mathcal{H},\text{ }\mathbf{l}^{i}=\delta ^{ii_{0}},\text{ }\mathbf{l=l}%
^{i}t^{i}=t^{i_{0}}  \label{a}
\end{eqnarray}%
where for the emergent pseudo-Goldstone modes $\mathbf{a}_{\mu }^{i}$ and $%
\mathbf{\phi }^{i}$ work the orthogonality conditions along an internal
symmetry breaking direction 
\begin{equation}
\mathbf{a}_{\mu }^{i}\mathbf{l}^{i}=\mathbf{\phi }^{i}\mathbf{l}^{i}=0\text{ 
}
\end{equation}%
\ One can readily see that the covariant derivative (\ref{55}) and (\ref{555}%
) in the Lagrangian $\mathcal{L}_{4D}$ (\ref{lag11}) generate some vector
field masses stemming from the first constant term in decomposition of the
effective Higgs field $\mathcal{H}$ in (\ref{deco}). Note that these mass
terms being proportional to the 5D Lorentz breaking mass scale $\mathrm{M}$
can not be gauged away since the restricted 4D gauge invariance (\ref{gt2})
is only left in the Lagrangian $\mathcal{L}_{4D}$ after compactification.
Meanwhile, this gauge invariance is spontaneously broken by itself, as
follows from the covariant derivative term (\ref{55}). Using a proper
unitary gauge one can decouple extra $\mathbf{\phi }$ scalar multiplet from
the 4-dimensional vector fields $\mathbf{a}_{\mu }^{i}$. As a result, they
are getting mass terms of the type%
\begin{eqnarray}
\mathcal{L}_{m}(\mathbf{a}_{\mu }) &=&\frac{1}{2}g^{2}\mathrm{M}^{2}Tr[%
\mathbf{a}_{\mu },\mathbf{l}]^{2}=\frac{1}{2}g^{2}\mathrm{M}^{2}\mathbf{a}%
_{\mu }^{i}\mathbf{a}_{\mu }^{j}Tr\{[t^{i_{0}},t^{i}]\cdot \lbrack
t^{i_{0}},t^{j}]\}  \notag \\
&=&\frac{1}{2}g^{2}\mathrm{M}^{2}\mathbf{a}_{\mu }^{i}\left(
f^{i_{0}}f^{i_{0}}\right) _{ij}\mathbf{a}_{\mu }^{j}  \label{ms}
\end{eqnarray}%
where the structure constants $f^{i_{0}ik}$ in the above commutators are
written in the matrix form $f_{ik}^{i_{0}}$ and matrix product $\left(
f^{i_{0}}f^{i_{0}}\right) _{ij}$ always appears diagonal for any internal
symmetry breaking direction $i_{0}$. It can easily be seen that these masses
crucially depend on this direction so that all the ground vector field modes
related to the corresponding broken generators of the internal symmetry
group $G$ receive the masses of the order of the 5D\ Lorentz scale $\mathrm{M%
}$. The masses vanish when there is a vanishing commutator $\left[
t^{i_{0}},t^{i}\right] =0$ in (\ref{ms}). This means that massless vector
bosons only occur when the index $i$ belongs to appropriate diagonal
subgroups of the symmetry group $G$.

Remarkably, the spontaneous breaking of internal symmetry also modifies the
masses of KK towers involved, as follows from the covariant derivative (\ref%
{555}). Indeed, properly writing out the commutators one comes to 
\begin{equation}
\overline{\mathbf{F}}_{\mu 5}^{s,i}=\partial _{\mu }\mathbf{A}%
_{5}^{s,i}-gf^{ijk}\mathbf{A}_{\mu }^{s,j}(\mathbf{\phi }^{k}+\mathrm{M}%
\mathbf{l}^{k})-i\frac{s}{R}\mathbf{A}_{\mu }^{s,i},\text{ }\mathbf{l}%
^{k}=\delta ^{ki_{0}}  \label{fs}
\end{equation}%
Unfortunately, in contrast to the previous case, we have no any conventional
gauge invariance to separate scalar and vector modes. Instead, one can
redefine the scalar fields in such a way to separate them in the momentum
space\footnote{%
Note that, though now one may not put an unitary gauge to get rid of the
massless scalar fields $\mathbf{A}_{5}^{s,i}$, these fields (particularly,
the Goldstone ones for $i$ values determined by non-zero matrix elements $%
f^{i_{0}ij}$ in (\ref{red})) correspond in fact to the unphysical particles
in the sense that they could not appear as incoming or outgoing lines in
Feynman graphs. In a somewhat similar context of the Standard Model
formulated in the axial gauge this was first argued in \cite{dams}.}

\begin{equation}
\mathbf{A}_{5}^{s,i}\longrightarrow \mathbf{A}_{5}^{s,i}\text{ }\mathbf{+}%
\text{ }ig\mathrm{M}\dfrac{k^{\mu }f^{i_{0}ij}\mathbf{A}_{\mu }^{s,j}}{k^{2}}
\label{red}
\end{equation}%
After their substitution into covariant derivative one has diagonalized
massless scalars $\mathbf{A}_{5}^{s,i}$ and massive vector towers $\mathbf{A}%
_{\mu }^{s,i}$ with modified kinetic terms determined by the scale $\mathrm{M%
}$ 
\begin{equation}
\left\vert \overline{\mathbf{F}}_{\mu 5}^{s}\right\vert ^{2}=\left( k_{\mu }%
\mathbf{A}_{5}^{s,i}\right) ^{2}+\left( g\mathrm{M}\right) ^{2}\left( g^{\mu
\nu }-\dfrac{k^{\mu }k^{\nu }}{k^{2}}\right) \mathbf{A}_{\mu }^{s,i}\left(
f^{i_{0}}f^{i_{0}}\right) _{ij}\mathbf{A}_{\mu }^{s,j}  \label{mass}
\end{equation}%
So, collecting the both types of mass terms for towers in (\ref{fs}) one has

\begin{equation}
\mathcal{L}_{m}(\mathbf{A}_{\mu }^{s})=\sum_{s=1}\mathbf{A}_{\mu }^{s,i}%
\left[ \frac{1}{2}\left( \frac{s}{R}\right) ^{2}\delta _{ij}+g^{2}\mathrm{M}%
^{2}\left( f^{i_{0}}f^{i_{0}}\right) _{ij}\right] \mathbf{A}_{\mu }^{s,j}
\label{zz}
\end{equation}%
We see that masses of towers for each number $s$ are significantly
influenced by an internal symmetry breaking along the direction determined
by the generator $\mathbf{l=l}^{i}t^{i}=t^{i_{0}}$. Particularly, all towers
related to the corresponding broken generators of the group $G$ will receive
the large extra masses of the order of the 5D Lorentz scale $\mathrm{M}$.

The most successful implementation of this phenomena may appear in Grand
Unified Theories considered ab initio in the five-dimensional spacetime.
Once the 5D SLIV is applied, along with the compactification to the physical
world, the adjoint "scalar field" multiplet composed from the extra vector
field components, $\left( \mathbf{A}_{5}^{0}\right) ^{i}$, will break GUT
down to the Standard Model so that all "non-diagonal" vector bosons (say, $X$%
- and $Y$- bosons in the $SU(5)$ theory) get large masses terms being of the
order of the scale $\mathrm{M}$. Thus, the scale $\mathrm{M}$ of the 5D SLIV
can be identified with the grand unification scale $M_{GUT}$ when emergent
GUTs are considered. This is in sharp contrast to the Abelian internal
symmetry case, where the 5D SLIV scale $M$ is arbitrary and could even be of
the order of a few \textrm{T}$\mathrm{eV}$.

The point is, however, that due to the high symmetry of the constraints (\ref%
{CONST}) one has in reality a vacuum degeneracy when applying them to the
internal symmetry breaking in the GUTs. Indeed, the first constraint in (\ref%
{CONST}) written for the proposed spacelike SLIV ($\mathbf{n}^{2}=-1$) as%
\begin{equation}
Tr\left( \mathbf{A}_{\overline{\mu }}^{0}\mathbf{A}^{0\overline{\mu }%
}\right) =\mathrm{M}^{2}\left[ 1+\mathcal{O}\left( (\mathbf{A}^{s})^{2}/%
\mathrm{M}^{2}\right) \right] \text{ ,}  \label{/}
\end{equation}%
where we also ignored all the higher KK modes, explicitly demonstrates such
a degeneracy in the internal space. Meanwhile, due to violation of the
starting gauge invariance (\ref{gt0}) in the post-compatification stage, the
radiative corrections will induce in general all possible potential terms in
the Lagrangian $\mathcal{L}_{4D}$ (\ref{lag11})%
\begin{equation}
\mathcal{U}(\mathbf{A})=\frac{m_{\mathbf{A}}^{2}}{2}\,Tr\left( \mathbf{A}_{%
\overline{\mu }}^{0}\mathbf{A}^{0\overline{\mu }}\right) +\frac{\lambda _{%
\mathbf{A}}}{4}[Tr\left( \mathbf{A}_{\overline{\mu }}^{0}\mathbf{A}^{0%
\overline{\mu }}\right) ]^{2}+\frac{\lambda _{\mathbf{A}}^{\prime }}{4}%
Tr\left( \mathbf{A}_{\overline{\mu }}^{0}\mathbf{A}^{0\overline{\mu }}%
\mathbf{A}_{\overline{\nu }}^{0}\mathbf{A}^{0\overline{\nu }}\right) \text{ }
\label{u}
\end{equation}%
where again the non-zero KK modes were omitted and some optional vector
field mass parameter ($m_{\mathbf{A}}^{2}$) and coupling constants ($\lambda
_{\mathbf{A}}$, $\lambda _{\mathbf{A}}^{\prime }$) introduced (higher order
terms are ignored). Now, one can readily see that the first two terms in the
potential $\mathcal{U}$ only add some constants to the Lagrangian because of
the constraint (\ref{/}), whereas the third one is in fact makes the lifting
vacuum degeneracy in a theory considered.

Let us turn again to the emergent $SU(5)$ GUT case. For the constraint (\ref%
{/}) and radiative corrections ignored there appears a twofold vacuum
degeneracy in the theory: one vacuum with $SU(3)\times SU(2)\times $ $U(1)$
symmetry (corresponding to the Standard Model) and another one with symmetry 
$SU(4)$ $\times $ $U(1)$. Interestingly, this resembles the vacuum
degeneracy problem in supersymmetric GUTs \cite{rabi1}. However, while there
is no way to split this vacuum degeneracy in the pure SUSY context, the
situation is radically changed in the emergent GUTs due to the radiative
corrections involved. Actually, one can readily confirm that for the
positive coupling constants $\lambda _{\mathbf{A}}^{\prime }$ in the
potential (\ref{u}) the $SU(3)\times SU(2)\times $ $U(1)$ vacuum is
definetely dominated in the emergent $SU(5)$\ theory. Remarkably, the
alternative $SU(4)$ $\times $ $U(1)$ vacuum may only exist for the negative
constants $\lambda _{\mathbf{A}}^{\prime }$, thus in an unstable theory
case, that is principally unacceptable. Although we do not calculate here
the above radiatively induced potential(\ref{u}), it seems natural to
propose that it may not destabilize the emergent $SU(5)$ theory, so that the
coupling constant $\lambda _{\mathbf{A}}^{\prime }$ always appears positive.
Thus, as a result of the degeneracy lifting, just the Standard Model vacuum
is generically chosen once the 5D SLIV occurs.

Due to radiative corrections, the $X$- and $Y$-bosons of $SU(5)$, apart from
the masses presented above in (\ref{ms}), receive extra mass contributions
(being proportional to $\lambda _{\mathbf{A}}^{\prime }\mathrm{M}^{2}$) from
the last term in the potential (\ref{u}). Likewise, all the diagonal
fifth-direction ground modes $\mathbf{A}_{5}^{0,i}=\mathbf{a}_{5}^{i}$ with
the $SU(3$)$\times SU(2)$ assignment $(8,1)+(1,3)+(1,1)$ receive the same
order masses $\mathcal{O}\left( \lambda _{\mathbf{A}}^{\prime }\mathrm{M}%
^{2}\right) $, while the non-diagonal Goldstone ones with the assignment $%
(3,2)+(\overline{3},2)$ appear massless though non-observable, as was
mentioned before$^{11}$. Analogously, an inclusion into the radiatively
induced potential (\ref{u}) the higher KK mode terms will produce masses for
still being massless fifth-direction $\mathbf{A}_{5}^{s,i}$ towers as well.
So, eventually all the starting 5D vector field modes, apart from the gauge
bosons of $SU(3)\times SU(2)\times $ $U(1)$, acquire masses in the 4D
emergent $SU(5)$\ GUT being automatically broken to the Standard Model.

\section{Conclusion}

We have argued that the spontaneously broken extra dimensional spacetime
symmetry could be a natural source for emergent vector bosons associated
with photons and other gauge fields. Indeed, the only way to produce such
bosons in a fully Lorentz covariant way would be to enlarge the existing
Minkowski spacetime to higher dimensions. As a matter of fact, all
four-dimension models only suggest a non-covariant description of vector
Goldstone bosons where one of vector field spacetime component $A_{\mu }$ ($%
\mu =0,1,2,3$) is inevitably "higgsified". Moreover, the spontaneous
breakdown of Lorentz symmetry itself may appear hidden from observation when
is considered in a gauge invariant theory framework.

The essential point is that an extra dimensional spacetime is eventually
reduced to a conventional four dimensions due to some compactification
pattern proposed. However, while the kinetic terms of the vector (and other)
fields\ will only possess a standard Lorentz symmetry after
compactification, their potential terms (or, equally, the polynomial vector
field constraints like (\ref{constr}) and (\ref{CONST})) may still have the
higher symmetrical form if the compactification pattern is properly chosen.
This consequently induce the high-dimensional SLIV due to which massless
pseudo-Goldstone states are generated as gauge boson candidates. So, an
adequate choice of a compactification mechanism is a crucial point when
considering extra dimensions as a possible source for a generation of
emergent gauge theories. However, while a simple compactification on a
circle conserves the starting spacetime symmetry for vector field
constraints like (\ref{constr}) and (\ref{CONST}), the orbifold
compactification $S_{1}/Z_{2}$ introduced to have chiral fermions in four
dimensions may in general explicitly break this symmetry down to a
conventional 4D Lorentz invariance.

Actually, for a conventional "asymmetrical" orbifold compactification when
ordinary four components of $A_{\overline{\mu }}$ are taken to be even under
the orbifold transformation, whereas its fifth component is odd, the 5D
Lorentz symmetry is turned out to be explicitly broken, though the 5D gauge
symmetry (\ref{ga1}) still remains. Eventually, one has the theory without
extra vector field components, $A_{5}^{0}$ and $A_{5}^{s}$ ($s=1,2,...$)
since the ground mode $A_{5}^{0}$ vanishes, while higher $A_{5}^{s}$ modes
appear absorbed by the 4D massive KK towers $A_{\mu }^{s}$. Without extra
vector field components, the nonlinear\ constraints (\ref{constr}) and (\ref%
{CONST}) will cause the VEV on one of ordinary components of the 4D vector
field ground mode $A_{\mu }^{0}$ and $\mathbf{A}_{\mu }^{0}$ ($\mu =0,1,2,3$%
), respectively. Thus, we come even in\ the 5D spacetime to the SLIV picture
appearing in the four-dimensional Nambu model and its generalizations (that
was intensively discussed above in section 1). Due to the starting 5D gauge
symmetry which really remains after compactification, the SLIV inducing
constraints (\ref{constr}) and (\ref{CONST}) will be simply converted into
the gauge fixing conditions so that such models have no observational
consequences unless this symmetry is explicitly broken by some external
sources.

In this connection, the 5D SLIV model developed above is entirely based on
the "symmetrical" orbifold compactification $S_{1}/Z_{2}$ under which all
spacetime components of the 5D vector field $A_{\overline{\mu }}(x,\theta )$
are taken to be even. Interestingly, such a "partially increased" Lorentz
invariance happens to significantly reduce an effective gauge symmetry
appearing for vector field components after compactification. The starting
gauge symmetry (\ref{ga1}) for KK states, both massive vectors $A_{\mu }^{s}$
($s=1,2,...$) and massless scalars $A_{5}^{s}$ ($s=0,1,2,...$), does not
work any longer. Only standard gauge invariance (\ref{ga2}) for massless
ground vector modes $A_{\mu }^{0}$ holds. This allows to uncover a number of
possible observational evidences in favor of emergent QED and Yang-Mills
theories which can not be gauged away as in 4D SLIV\ theories. They include,
apart from conventional gauge couplings, the properly suppressed direct
multi-photon (multi-boson, in general) interactions. This means that
emergent gauge theories actually possess only a partial gauge invariance
whose observation could shed light on their high-dimensional nature. Another
crucial prediction is an existence of the unabsorbed fifth-direction
non-zero modes $A_{5}^{s}$ ($s=1,2,...$) being massless at the tree-level.
Due to KK parity they can be only produced by pairs from an ordinary matter
being properly suppressed by the 5D Lorentz violation scale $\mathrm{M}$ or
by the compactification mass $M_{KK}\sim 1/R$. On the other hand, any heavy
KK state will now rapidly decay into the $A_{5}^{s}$ mode plus ordinary
matter that seems to invalidate the dark matter scenario related to extra
dimension \cite{hoop}. However, these $A_{5}^{s}$ modes being no more
protected by gauge invariance could in principle acquire large masses
through radiative corrections so that the lightest KK state may appear
rather stable to provide the measured dark matter density.

All the above, while was largely spoken relative to the emergent QED, is
equally applicable to both Abelian and non-Abelian cases. However, there is
one point being particularly specific to Yang-Mills theory. In this case,
due to 5D SLIV, together with the spacetime symmetry breaking, the
non-Abelian internal symmetry group $G$ also occurs spontaneously broken. As
a result, all "non-diagonal" emergent vector bosons appearing during
symmetry breaking process (\ref{so}) may acquire masses. This breaking
origins from the extra vector field components playing a role of some
adjoint scalar field multiplet in the 4D spacetime. Therefore, one may have
the generic Higgs effect in the 5D SLIV theory which breaks the starting
internal symmetry $G$ to its diagonal subgroups. When being applied to Grand
Unified Theories this results in a fact that the emergent GUTs automatically
appear broken down to the Standard Model just at the 5D Lorentz violation
scale $\mathrm{M}$. So, a spontaneous breakdown of a high-dimensional
spacetime symmetry to a conventional Lorentz invariance may determine an
internal symmetry pattern at low energies, and also control an admissible
proton decay rate and, consequently, an acceptable matter-antimatter
asymmetry in the early universe. We may return to this interesting scenario
elsewhere.

\section*{ Acknowledgments}

One of us (JLC) thanks Colin Froggatt, Rabi Mohapatra and Holger Nielsen for
interesting discussions and correspondence. This work is partially supported
by Georgian National Science Foundation (Contracts No. 31/89 and No.
DI/12/6-200/13).

\end{document}